# How much research shared on Facebook happens outside of public pages and groups?

A comparison of public and private online activity around PLOS ONE papers

Asura Enkhbayar[1], Stefanie Haustein[1,2,3], Germana Barata[4], Juan Pablo Alperin[1,5,*]

[1] Scholarly Communications Lab, Simon Fraser University, Vancouver (Canada)
[2] School of Information Studies, University of Ottawa, Ottawa (Canada)
[3] Centre interuniversitaire de recherche sur la science et des technologies (CIRST), Université du Québec à Montréal, Montréal (Canada)
[4] Laboratory of Advanced Studies in Journalism, State University of Campinas (Brazil)
[5] School of Publishing, Simon Fraser University, Vancouver (Canada)
*Corresponding author: juan@alperin.ca

Despite its undisputed position as the biggest social media platform, Facebook has never entered the main stage of altmetrics research. In this study, we argue that the lack of attention by altmetrics researchers is due, in part, to the challenges in collecting Facebook data regarding activity that takes place outside of public pages and groups. We present a new method of collecting aggregate counts of shares, reactions, and comments across the platform—including users' personal timelines—and use it to gather data for all articles published between 2015 to 2017 in the journal PLOS ONE. We compare the gathered data with altmetrics collected and aggregated by Altmetric. The results show that 58.7% of papers shared on Facebook happen outside of public spaces and that, when collecting all shares, the volume of activity approximates patterns of engagement previously only observed for Twitter. Both results suggest that the role and impact of Facebook as a medium for science and scholarly communication has been underestimated. Furthermore, they emphasise the importance of openness and transparency around the collection and aggregation of altmetrics.

*Keywords: altmetrics; Facebook; public engagement; science communication*

## 1.0   Introduction

As of October 1, 2019 Facebook reports 2.49 billion monthly active users (Facebook, 2019)—approximately 32.3% of the world's population (United Nations, 2019). The rise of Facebook has been accompanied by the rapid "platformization" of the web in general (Helmond, 2015), and academia has been no exception (Chan, 2019). Like more academically-oriented sites (i.e., Academia.edu and ResearchGate), Facebook has become an important tool as well as subject of research. As the extent of platformization and its impact on society receives increased attention (Zuboff, 2015), several studies have shown that the platform is being used in an educational context, outside and inside of classrooms (Roblyer, McDaniel, Webb, Herman, & Witty, 2010; Selwyn, 2009). The general public uses Facebook mostly for purposes of information sharing (Baek, Holton, Harp, & Yaschur, 2011). In academia, almost 40% of scientists and engineers use it regularly, as do 50% of scholars in the social sciences, arts, and humanities (McClain, 2017; Van Noorden, 2014).

At the same time, the growing importance of social media as a medium for scholarly communication has led to the creation and consolidation of a new field of study known as altmetrics, which tries to quantify and understand how research circulates online, including on social media platforms (Erdt, Nagarajan, Sin, & Theng, 2016). Yet, despite the increasing number of altmetric studies and its social significance and use as a platform for sharing information, there has been limited research that specifically focuses on the circulation of research on Facebook (Enkhbayar & Alperin, 2018; Freeman, Roy, Fattoruso, & Alhoori, 2019; Ringelhan, Wollersheim, & Welpe, 2015).

One reason for the lack of research might be that, according to most altmetric studies, research does not appear to be shared on Facebook very broadly. In their review of nine studies, Erdt et al. (2016) report that only 7.7% of research articles are shared on Facebook —far less than 59.2% that are found for Mendeley libraries or 24.3% for Twitter. The coverage can vary depending on discipline—with biomedical and health sciences among the areas with the highest coverage (Costas, Zahedi, & Wouters, 2015; Fenner, 2013)—and country, with studies written by authors in Brazil being shared most among Latin American journals (Alperin, 2015). Additionally, the coverage has been reported to vary by data aggregators. For example, Twitter coverage for PLOS ONE publications was reported between 23.9% on Plum Analytics and 57.0% on Altmetric. Similarly, the coverage for Facebook fluctuated between 7.9% on Lagotto and 16.3% on Plum Analytics (Zahedi & Costas, 2018b). Overall, Facebook coverage has been reported to be significantly lower than the ones observed for Twitter (Haustein, Costas, & Larivière, 2015; Xia et al., 2016; Zahedi & Costas, 2018b), which corroborates findings that academics prefer Twitter over Facebook for science outreach (Hassan et al., 2017; McClain, 2017; Thelwall, Haustein, Larivière, & Sugimoto, 2013).

There is also significant variation depending on the way in which the mentions of research on Facebook are retrieved and aggregated (Zahedi & Costas, 2018a). Searching for mentions of research by querying Facebook's API, which measures the number of times links are shared anywhere on Facebook, identifies more mentions than when using data mining approaches, which can only be performed on public Facebook pages. The latter approach is used by Altmetric, while Plum Analytics makes use of the API. As we argue here, although quite technical, this methodological distinction has far-reaching implications for results of commonly used altmetric indicators and thus the understanding of the role of Facebook in the circulation of research. Analyzing and comparing retrieval methods is thus a central issue with regards to the reliability and reproducibility of altmetrics and, as a consequence, for the field of altmetrics in general (Haustein, 2016).

More specifically, the limitation of working with public-only pages may omit a significant portion of the research-related activity that happens on the platform. A recent study shows that academics on Facebook tend to share science on the platform in a personal and intimate manner, rather than professionally for science outreach and communication (McClain, 2017). This would suggest that looking only at public pages, like is done by the majority of altmetric studies, misses a significant amount of Facebook activity related to scholarly papers. McClain (2017) also suggest differences in how academics view and use public and private spaces within social media platforms such as Facebook. As such, in this study we seek to measure the extent and nature of this difference between public and private acts related to scholarly documents (Haustein, Bowman, & Costas, 2016), with *the goal of understanding how much engagement with research is taking place on Facebook, including both public and private spaces.*

## 2.0   Background

### 2.1   Methods for collecting Facebook altmetrics

There are two general approaches used by those seeking to collect metrics related to the circulation of research on Facebook: The first is the extraction of mentions of articles from the text of posts in a curated list of public groups and pages, presumably through data-mining approaches. This method is used by Altmetric[1]. The second is to use Facebook's API to query engagement counts, as calculated by Facebook. In contrast to the first approach, the API approach covers only counts but of activity across all of Facebook, including engagement that takes place in user's private pages. PlumX Metrics[2] and the method developed for this study use the API approach.

#### *2.1.1   Extracting mentions from public pages on Facebook*

A data-mining approach is able to capture the number of mentions an article receives in posts on a curated list of public Facebook pages and groups. This approach is used by Altmetric since October 2011 (Altmetric, 2019a). To do this successfully, they use a list of public pages and groups that can then be monitored for posts that contain basic metadata about the article in question. Although Altmetric declined to provide specific details of their text processing methods, they do report that they use the "Pages" endpoint of the Facebook's Graph API and that they monitor known web domains (i.e., the first part of a URL), including links that have been "shortened" (Altmetric, 2019b). A review of posts identified by them also reveals that they additionally monitor some identifiers (e.g., DOIs), but, although it is theoretically possible, we found no evidence that they monitor additional metadata (e.g., title or author names). Moreover, they attempt to identify and combine the mentions of research outputs with several major identifiers (e.g., the DOI, PubMedID, arXiv ID, ADS ID, SSRN ID, RePEC ID and ISBN) into a single record. It is not clear in which ways Altmetric identifies multiple URL variations for the same article (e.g., links to other output formats, such as a PDF), or what pages are included in their curated list.

One of the main strengths of this approach is that the resulting data can be linked to a page that is publicly available, making such data auditable (e.g., it is possible to see the full content of the Facebook post to verify the existence of the link to the research, as well as the identity of the person who posted it). The major limitation, however, is that it can only be done for posts that are on public pages, as Facebook does not share—and so Altmetric cannot collect—the texts of posts that are private.

#### *2.1.2   Querying the Graph API for engagement counts*

The second approach makes use of Facebook's Graph API to access engagement counts for what Facebook calls *objects* (i.e., groups of URLs that Facebook has determined to refer to the same content) in their *social graph* (i.e., the network of users that make up Facebook's content). These engagement counts comprise the number of shares (i.e., URL posted by a user), reactions (i.e., 'likes' or other emotion icons), comments (i.e., responses to a shared URL), and plugin comments (i.e., comments created by users on an external URL using the Facebook comments plugin).

---
1 https://www.altmetric.com/
2 https://plumanalytics.com/learn/about-metrics/

Facebook's Graph API accepts a single URL as a parameter, and returns both a Facebook Object ID, along with the number of times the object has been shared, liked, and commented on by users across Facebook. This includes posts on users' Timelines (largely restricted from public view) as well as posts on public pages (visible to anyone).[3]

To the best of our knowledge, since Lagotto stopped to collect Facebook metrics due to changes to the Graph API (Fenner, 2014) and PlumX Metrics replaced individual counts for shares, likes, and comments with the total count of all actions (Allen, 2016), no altmetric provider presents Facebook counts across platforms and identifiers using the API method. There are, however, plans in place by Our Research (formerly ImpactStory) and Public Knowledge Project, in collaboration with Crossref, to add Facebook as a data source for Crossref's Event Data (Alperin, Enkhbayar, Piwowar, Priem, & Wass, 2018).

One of the main strengths of this approach is that it can access activity around research articles that happen outside of public view (at the expense of allowing that activity to be audited). A limitation is that the data collected are not auditable, and so while it is possible to track a larger volume of engagement counts, it is not possible to view the underlying posts that led to it taking place. Another limitation is that due to the reliance on specific URLs for querying, this method is particularly sensitive to what URLs are identified for an article. In previous work, we built on Wass' (2018) detailed overview of the challenges of working with DOIs and URLs and subsequently with Facebook's Graph API (Enkhbayar & Alperin, 2018). In this study, we explore the relation of an *external URL* to the Facebook internal *Open Graph objects* and *engagement counts* that is foundational to any approach that collects Facebook metrics for scientific articles using their API. We also show how the API-based approach can be used to combine the results from multiple URLs that may exist for the same article, as well as those related to various major identifiers (e.g., DOIs and PMIDs). In Figure 1 we show an idealized case of an article with several related URLs and identifiers whose Facebook Object IDs can then be used to arrive at a single aggregated total engagement count.

---

[3] Although this method collects information about activity that happens outside of public view, it relies solely on available data that is provided and displayed by Facebook itself. None of the collected data contains any information that can be used to identify individuals.

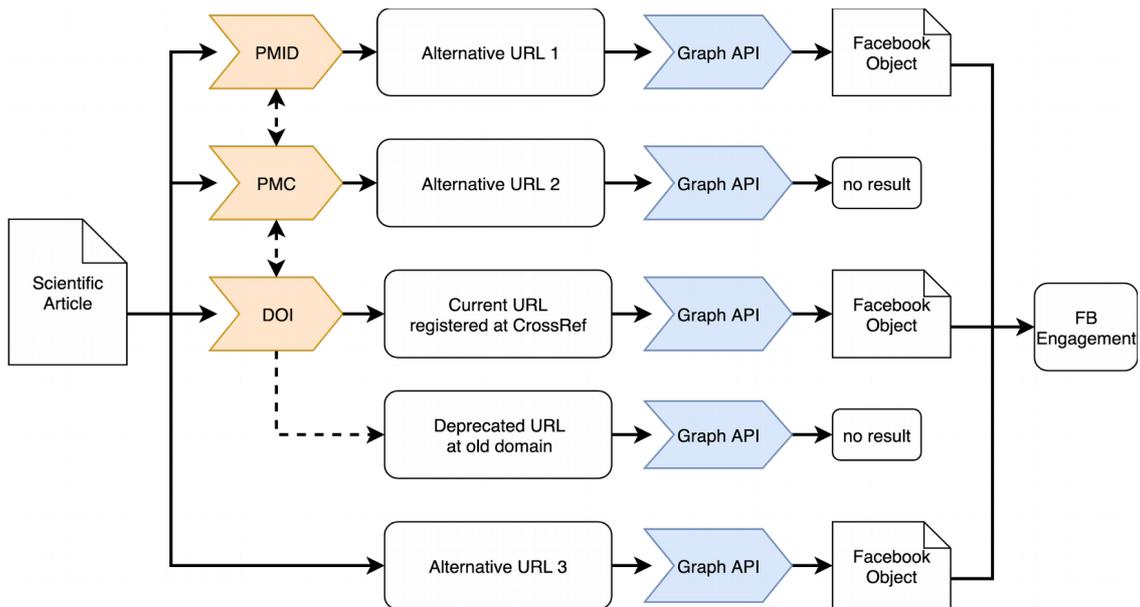

Figure 1. Idealized case for the collection of Facebook metrics for a scientific article using multiple URLs

## 2.2 Differences between collection methods

Both approaches are limited to some extent as they rely on knowing all of the URLs and identifiers that can be used to refer to a given research article. However, there are two important differences between both methods:

1. *Type of engagement*. The number of "mentions" by means of textual analysis of a post has a different conceptual meaning than the count of shares, reactions, or comments as provided by Facebook's API. Furthermore, some data providers (e.g., Plum X) decide to aggregate the reported number of shares, likes, and comments to provide a single "Facebook count". Thus, it is important to differentiate between reported Facebook counts which could represent one of several things: mentions on public posts, one of several reported Facebook metrics (shares, reactions, comments), or the sum of these metrics.

    In this paper, we primarily discuss and compare the Facebook *shares* we retrieved with our version of the second method with the *mentions captured by* Altmetric using the first. We are unable to use Plum's implementation of the API approach because they do not make data available for research We also compare both of these to tweets and retweets on Twitter (as collected by Altmetric), which are considered similar forms of *appraisal acts* to posting links on Facebook (Haustein et al., 2016).

2. *Scope of engagement*. Both approaches cover different scopes of content available on Facebook. In order to access the full text of Facebook posts, they need to be posted in public groups and pages that are tracked by the data aggregator. In contrast, querying Facebook's API provides a way to access the counts across all of Facebook's posts including user posts on each other's Timelines, those made in private or closed groups, as well as in any other group, regardless of whether it has been previously identified as being relevant.

In this paper, we will refer to mentions collected using the first method—used to analyze public posts—as *public-only shares* (POS) and mentions collected using the second method—used to analyze posts anywhere on Facebook—as *all-engagement.* When this engagement refers specifically to the sharing of a link on Facebook (as opposed to comments or reactions to that link), we refer to it as *all-engagement shares* (AES)*.*

A summary of the similarities and differences between both approaches can be found in Table 1.

Table 1. Comparison of *public-only* and *all-engagement* metrics for Facebook

| **Method 1: Public-only metrics** | **Method 2: All-engagement metrics** |
|---|---|
| ⊖ Only captures number of mentions | ⊕ Provides several metrics: number of shares, sum of all reactions, comments, and comments made outside of Facebook using Facebook plugins |
| ⊖ Only captures posts on public pages | ⊕ Captures any activity across Facebook |
| ⊕ Captures occurrences of any link variant for an article at known publisher domains | ⊖ Only captures specific URLs that are queried |
| ⊕ Captured engagement is auditable (posts with mentions and their authors can be reviewed) | ⊖ Reported engagement numbers are aggregated for all users and can't be reviewed or disambiguated |
| ○ Limited by the curation of relevant identifiers and URLs | ○ Limited by the curation of relevant identifiers and URLs |
| ○ Data provided by Altmetric API for a publication by using one of several persistent identifiers (access key required) | ○ Data accessed through Facebook's Graph API on URL level (developer key required) |

## 2.3  How do studies address these differences?

Both approaches have been studied to some extent in the existing literature. Studies which compare altmetrics to citations and to each other often rely on data provided by Altmetric (e.g., Costas et al., 2015; Haustein et al., 2015; Thelwall et al., 2013). In contrast, studies that have sought to compare several data aggregators are accordingly comparing different collection methods depending on the choice of data providers (Zahedi & Costas, 2018a; Zahedi, Fenner, & Costas, 2014).

Stark differences can be observed in the degree and clarity in which the differences between collection methods for Facebook are addressed. A few publications did neither address the *type of engagement* nor the *scope of engagement* collected (Peters, Kraker, Lex, Gumpenberger, & Gorraiz, 2015; Robinson-García, Torres-Salinas, Zahedi, & Costas, 2014; Xia et al., 2016; Zahedi & Costas, 2018a). Half of these studies only used a single data provider, while the other compared the data from several providers.

The vast majority of studies that investigated Facebook as an altmetric source did, either implicitly or explicitly, specify the *type of engagement* represented in their data (e.g.

Chamberlain, 2013; Fenner & Lin, 2014; Freeman et al., 2019; Hassan et al., 2017; Priem, Groth, & Taraborelli, 2012; Ringelhan et al., 2015; Thelwall et al., 2013). However, the *scope of engagement* would still be disregarded in the methods sections.

Starting in 2014, several publications have not only clearly addressed the *type of engagement*, but also specified the *scope of engagement* for collected Facebook metrics (Haustein, 2016; Haustein et al., 2015; Zahedi & Costas, 2018b; Zahedi et al., 2014; Zahedi, Fenner, & Costas, 2015).

Overall, the scope of captured engagement seems to be overlooked more frequently than the type of captured engagement, with the remark that studies comparing several collection methods were more likely to mention the scope of engagement. With the increasing importance of Facebook as a social platform used by academics in professional as well as private settings, revisiting the role and impact of Facebook as a data source for altmetric research might be required. As part of that undertaking, we suggest to pay closer attention to the difference between *public-only share* (POS) and *all-engagement share* (AES) counts as collected (and compared) by various data aggregators.

## 3.0   Methodology

To ensure that all articles received the same opportunity to be shared on Facebook, avoiding journal biases and differences between paywalled and open access articles, while at the same time maintaining a sufficiently large sample, we analyze all articles published in the journal PLOS ONE between 2015 and 2017. As a multidisciplinary and open access mega journal, PLOS ONE provides the advantage of publishing a wide variety of fields of research free to read and accessible by the general public.

Using the *rplos package* for R (Chamberlain, Boettiger, & Ram, 2018) on July 17th, 2018, we retrieved a total of 61,872 articles that were published in PLOS ONE from the beginning of 2015 until the end of 2017. The precise query used can be found in box 1.

```
searchplos(q="*:*",
  fl="id, publication_date, title, author",
  fq=list('publication_date:[2015-01-01T00:00:00Z TO 2017-12-31T23:59:59Z]',
   'journal_key:PLoSONE',
   'doc_type:full'))
```
Box 1. Query used to retrieve PLOS publications using *rplos* package.

We used the DOIs of these 61,872 documents to query the Altmetric API and retrieved 50,819 responses with at least one mention across all of the sources tracked by Altmetric. Responses from the Altmetric API include mentions to various versions of the article, including those in PubMed, PubMed Central, and arXiv, when available. Of these, 9,632 (19.0%) had at least one Facebook mention and 43,083 (84.8%) had at least one Twitter mention. These correspond to our *POS* and *Twitter (TW)* datasets respectively.

Collecting the *AES* data was more complex. First, we identified ten different URL patterns that users might have used to share the respective article on Facebook (Table 1). The first two URLs are based on Crossref's guidelines for creating links from a DOI. In March 2017,

Crossref changed the recommended DOI link from http://dx.doi.org/{doi} to https://doi.org{doi} (Hendricks, 2017). URLs 3 to 8 correspond to various links that can be found on the PLOS ONE website itself, including the landing page URL, links to various sub-pages (i.e., authors, comments, related articles, metrics) and the link to the PDF version of the article. Additionally, for each DOI, we attempted to retrieve the PubMed ID and Pubmed Central ID with the identifier converter provided by the National Centre for Biotechnology Information (NCBI).[4] Only five articles (two corrections and three announcements by the PLOS staff) did not return a Pubmed ID. These articles have corresponding landing page URLs for PubMed and PubMed Central (URLs 8 and 9) that incorporate the respective IDs if found.

Table 2: Overview of URL types based on three identifiers (DOI, Pubmed ID, Pubmed Central ID)

|    | URL type | Pattern |
| --- | --- | --- |
| 1 | doi | https://doi.org/{**doi**} |
| 2 | doi_old | http://dx.doi.org/{**doi**} |
| 3 | landing | http://journals.plos.org/plosone/article?id={**doi**} |
| 4 | authors | http://journals.plos.org/plosone/article/authors?id={**doi**} |
| 5 | metrics | http://journals.plos.org/plosone/article/metrics?id={**doi**} |
| 6 | comments | http://journals.plos.org/plosone/article/comments?id={**doi**} |
| 7 | related | http://journals.plos.org/plosone/article/related?id={**doi**} |
| 8 | pdf | http://journals.plos.org/plosone/article/file?id={**doi**}&type=printable |
| 9 | pubmed | https://ncbi.nlm.nih.gov/pubmed/{**pmid**} |
| 10 | pmc | https://ncbi.nlm.nih.gov/pmc/articles/{**pmcid**}/ |

We then used the URL endpoint of the Graph API v2.10[5] to retrieve the *AES* counts for each of the 618,720 URLs. We started the data collection script on July 18, 2018 and finished July 23, 2018. Of the 618,720 URLs queried, a Facebook Object was found for 69,983 (11.3%) of cases. However, due to the way Facebook's API works, 37,062 of the 69,983 URLs (53.0%) were mapped to Facebook Objects with zero shares, reactions, and comments. These were treated the same as those which were not found in the API.

Each of the remaining Facebook Objects were stored, along with their Object ID. Facebook attempts to assign the same Object ID to URLs that correspond to the same content, such as those that use different protocols (i.e., *https* vs *http*), those that redirect to the same page (e.g., links with or without a *www* at the beginning), or those that have additional characters at the end (e.g., a "trailing slash" or some URL parameters).

In a previous study (Enkhbayar & Alperin, 2018), we analyzed a random sample of 100,000 articles to investigate the mapping between DOIs, their respective URLs and Facebook Objects—a challenge that has been described and studied in parts by Wass (2016). Although best practices by publishers (such as the proper use of 'canonical URLs' and metatags) improve the mapping, using a random sample of over 100,000 DOIs and only 4

---

4 https://www.ncbi.nlm.nih.gov/pmc/tools/id-converter-api/
5 https://developers.facebook.com/docs/graph-api/reference/v2.10/url

URL variants for each article, we encountered problems in the mapping between URLs and Facebook Objects in 12% of cases. Using the same methods, we could not unambiguously map a Facebook Object to an article for 24 (0.04%) articles in our PLOS dataset. The reason for the difference between the degree of discrepancies can be traced to the higher homogeneity of this dataset. All articles in this study have been published within a timeframe of three years and at the same journal, whereas the previous study analyzed a random sample of 100,000 articles indexed in the Web of Science, published between 2009 and 2015 (Piwowar et al., 2018). Thus, after removing all problematic articles the final dataset contains 61,848 articles.

Because of the interdisciplinary nature of PLOS ONE, we used the article-level subject classifications from Piwowar et al. (2018) to assign each article to a single discipline by assigning it to its most frequently cited NSF specialty (as indexed by the Web of Science). The NSF classification system used in Piwowar et al. (2018) comes in three levels of granularity: Grand disciplines, disciplines, and specialities. In our analysis we focus on the disciplinary level as it provides a good midway point between a big picture and detail. However, specialty classifications are used in times when a detailed look seems appropriate and useful as in the case of reported correlations.

The disciplines of Arts (2 articles) and Humanities (15) are excluded from the analysis due to the low number of publications. By matching DOIs and titles between the PLOS publications that we collected and were provided for the disciplinary analysis we arrived at 57,902 articles with a grand discipline, discipline, and specialty, while 3,929 articles are missing those. Of these 3,929 articles without disciplinary information, 3,374 articles were either corrections, retractions, or published by the PLOS ONE staff which are not considered traditional research outputs and, consequently, not included in the data provided by Piwowar et al. (2018). Thus, only 555 articles are missing disciplinary information due to miscellaneous reasons (e.g., errors in the metadata). These 3,929 articles without disciplinary information are excluded from analyses pertaining to disciplines, but not dropped from the overall dataset.

**Openness and reproducibility.** The data used to produce results is available at [doi:10.7910/DVN/3CS5ES](doi:10.7910/DVN/3CS5ES) (Enkhbayar, Haustein, & Alperin, 2019) and has been published under a CC0 license. The code to reproduce all figures and tables is available at [doi:10.5281/zenodo.3381821](doi:10.5281/zenodo.3381821) (Enkhbayar, 2019) and is published under the MIT license. The code to collect the original data can be found at [doi:10.5281/zenodo.1314990](doi:10.5281/zenodo.1314990).

## 4.0   Results

We organize our results in two broader areas: An analysis of coverage for the different collection methods and an analysis of the volume of engagement found. To address the prior, we present the coverage for all metrics (AES, POS, TW) followed by a detailed look at the difference between the two Facebook methods, including a disciplinary breakdown. To address the latter, we compare the extent of engagement across the three metrics, followed by a detailed inspection of the Facebook share counts, again including a disciplinary breakdown.

## 4.1 Comparison of retrieval methods

We compare AES retrieved from the Graph API using our method with two metrics collected from Altmetric: POS on Facebook and the number of TW on Twitter. As previously discussed, Twitter is considered to be the most used social media platform for disseminating research, and as such serves as a useful point of comparison.

### 4.1.1 Coverage

#### Comparison of AES, POS, and TW

Table 3 shows the coverage for AES, POS, and TW across all years for all 61,848 articles. Twitter shows the highest coverage for each year, covering 43,064 (69.6%) articles in total. These coverage numbers are followed by the coverage determined by the Graph API method totalling 34.6% (21,415) articles. Finally, Altmetric's POS displays the lowest coverage among these three metrics with 15.6% (9,623) articles found for all years. The proportion of articles covered across individual years remains stable for each method.

Table 3. Coverage for Twitter (TW), all-engagement shares (AES), and public-only shares (POS).

|           | AES            | POS           | TW              | All articles |
|-----------|----------------|---------------|-----------------|--------------|
| 2015      | 8,596 (33.8%)  | 3,981 (15.7%) | 16,976 (66.8%)  | 25,427       |
| 2016      | 6,992 (37.2%)  | 3,058 (16.3%) | 13,807 (73.4%)  | 18,809       |
| 2017      | 5,827 (33.1%)  | 2,584 (14.7%) | 12,281 (69.7%)  | 17,612       |
| All years | 21,415 (34.6%) | 9,623 (15.6%) | 43,064 (69.6%)  | 61,848       |

Of the 46,286 articles that are covered by at least one of the three methods, 15.7% (7,270) articles are covered by all three data collection method, while half (49.6%, 22,964) articles were only shared on Twitter. One quarter (24.3%, 11,226) publications were both tweeted and shared on Facebook but have not been indexed as POS by Altmetric. Overall, only 3,222 (7.0%) articles shared on Facebook were not shared on Twitter but were caught by one of the two Facebook collection methods. Figure 2 displays the exact counts and overlaps of each method.

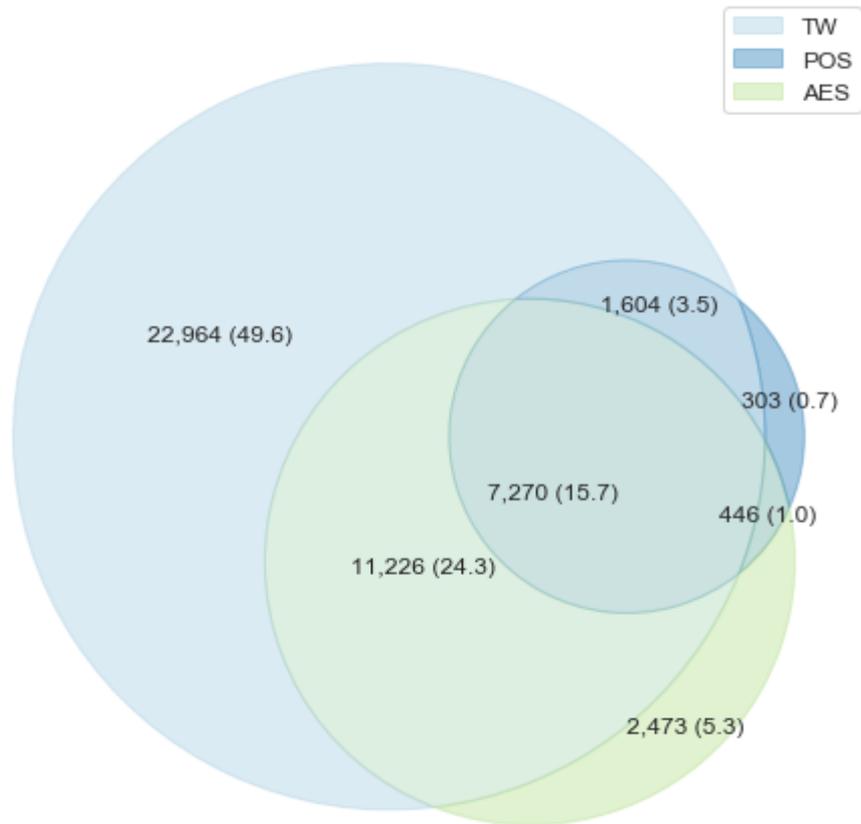

Figure 2. Coverage of articles that received one of the three engagement types (n=46,286) for Twitter (TW), public-only shares (POS), and all-engagement shares (AES).

Facebook coverage in detail

Focussing on the Facebook data collection methods shows that the proportion of articles found with each method remains constant across years (Table 4). On average, of all articles that were shared on Facebook (i.e., found in AES and/or POS; n=23,322) 7,716 articles (33.1%) are shared by both approaches. 13,699 (58.7%) are only found with the API method, while 1,907 articles (8.2%) are only found with Altmetric's approach.

Table 4. Coverage of articles that were only found by all-engagement shares (AES) and public-only shares (POS) respectively, in addition to articles that were found by both approaches. The number of articles found by either articles is displayed in the last column.

|           | AES            | AES and POS    | POS           | Any FB |
|-----------|----------------|----------------|---------------|--------|
| 2015      | 5,414 (57.6%)  | 3,182 (33.9%)  | 799 (8.5%)    | 9,395  |
| 2016      | 4,540 (59.8%)  | 2,452 (32.3%)  | 606 (8.0%)    | 7,598  |
| 2017      | 3,745 (59.2%)  | 2,082 (32.9%)  | 502 (7.9%)    | 6,329  |
| All years | 13,699 (58.7%) | 7,716 (33.1%)  | 1,907 (8.2%)  | 23,322 |

Coverage by disciplines

Half of all PLOS ONE output is classified as Clinical Medicine (50.7%), one quarter as Biomedical Research (24.8%) and 11.7% as Biology (Figure 3). Psychology (2.9%), Engineering and Technology (2.0%), Earth and Space (1.9%), Health (1.9%), Social Sciences (1.3%) and Physics (1.1%) follow as most frequent disciplines published in PLOS ONE. Twitter (TW) consistently shows a higher coverage for every discipline, ranging from the highest in Psychology, where almost every paper was distributed on Twitter (94.2%, 1,583), to the lowest in Chemistry (44.4%, 150). Even the lowest Twitter coverage is still higher than the highest covered discipline for *POS* (Social Sciences: 35.0%, 256). AES range from the highest coverage in the Social Sciences (70.5%, 516) to Chemistry (8.0%, 83). The distribution of disciplines for all articles and for the different collection methods (AES, TW, and POS) and be found in Figure 3 and in Table form in appendix A.

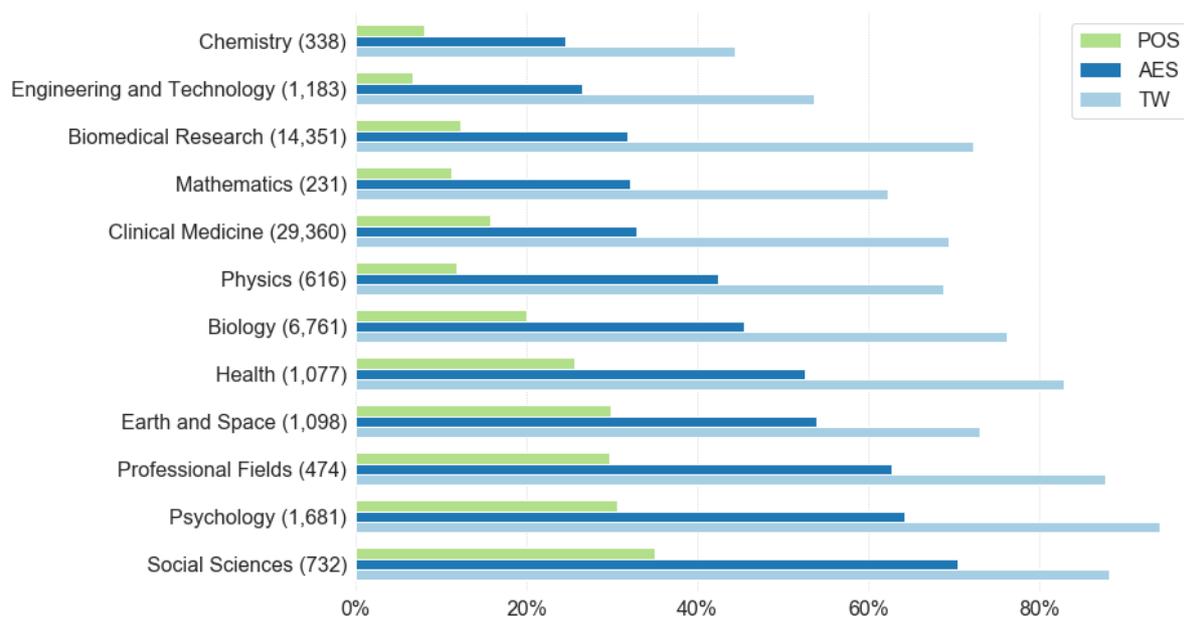

Figure 3. Coverage of disciplines for AES, POS, and TW.

Difference between Facebook methods per discipline

Combining retrieval methods, a total of 22,928 articles were shared on Facebook (i.e., found in AES and/or POS). One third of these articles (7,612; 33.2%) were found by both approaches, more than half (13,464; 58.7%) were only found by the Graph API method, while Altmetric retrieves 8.1% (1,852) of articles which are not found via the Graph API. This means that 41.3% of Facebook activity related to PLOS ONE papers is shared in public spaces, while 58.7% are shared in exchanges that are not captured by POS approaches. Appendix B provides a full breakdown of the share of articles that were found privately, publicly, or by both methods across all disciplines.

The ratio of posts on public pages (articles found by both methods vs. only POS) and all posts (articles found by both and only AES) is highest in Earth and Space (55.1%), followed by Social Sciences (49.6%), Health (48.7%), Psychology (47.9%) and Professional Fields (47.5%) (Figure 4). In these fields, almost half of all Facebook engagement is in public spaces. On the other end of the spectrum, in Engineering and Technology (24.8%), Physics (28.0%), Chemistry (32.5%) and Mathematics (35.1%), the great majority of all Facebook activity happens outside of public pages and groups and is thus also not captured by Altmetric.

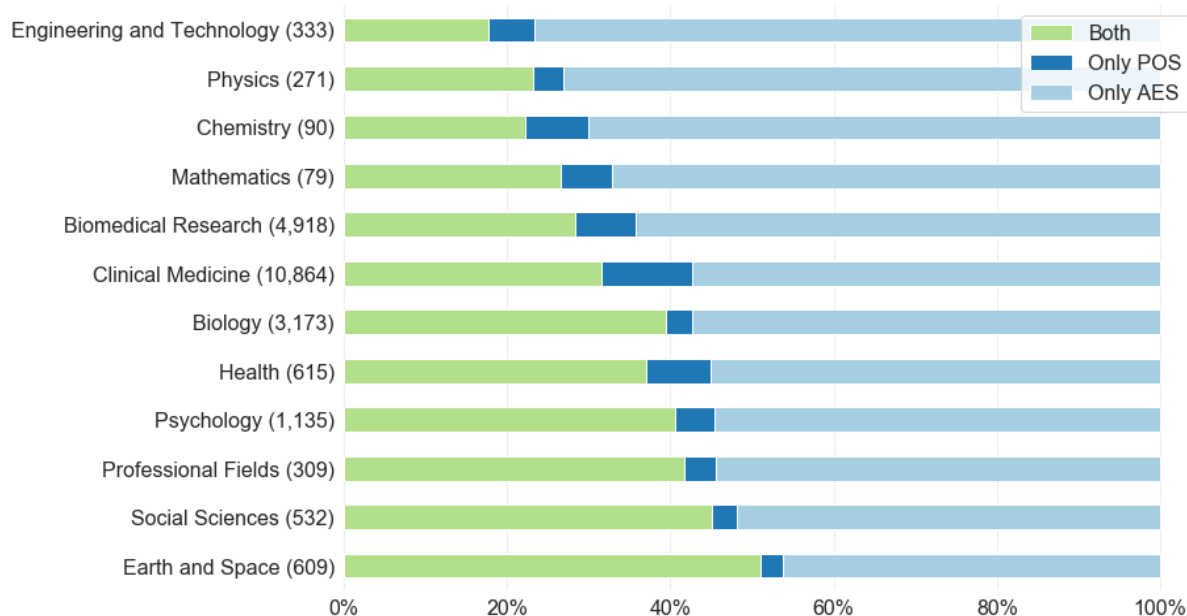

Figure 4. Percentage of papers found by both Facebook retrieval methods, public-only shares (POS) via Altmetric only and all-engagement shares (AES) via Facebook API only per NSF discipline. Articles found by both methods and POS only are assumed to reflect public, those found by AES to show private activity.

Despite missing 58.7% of articles overall, Altmetric's approach indexes mentions for a significant number of articles that are not captured by the Graph API (Figure 4). Clinical Medicine stands out with the highest number of articles that were only indexed by Altmetric's method (11.0%), while Earth and Space (2.6%), Social Sciences (3.0%), and Biology (3.3%) display the lowest shares. In total, Altmetric found 1,853 (8.1%) articles that couldn't be indexed by our method.

### 4.2    Engagement counts

Having compared coverage overall and at the discipline level, we inspect the volume of engagement (i.e., counts) that we retrieved (AES) with Altmetric's POS and TW.

#### Comparison of AES, POS, and TW

We calculated basic descriptive statistics for all three metrics (Table 5). POS display less extreme maximum engagement counts, as well as a lower median and geometric mean[6] compared to both AES and TW. While AES report the highest maximum engagement count in the data, both median and geometric mean show that TW is still, on average, registering higher engagement counts. In addition to comparing the distributions for each metric, we calculated Spearman correlations between each pair, after replacing missing values with zeros (Table 6).[7] Correlations were highest between AES and POS ($\rho$=.48), followed by AES and TW ($\rho$=.45) and POS and TW ($\rho$=.36).

---

6 The geometric mean is the calculation of the arithmetic mean based on log-transformed values, followed by a reverse log-transformation to the original scale. Thus, this approach is less prone to extreme values in the data and more appropriate for the given dataset.

Table 5. Descriptive statistics for each distribution including the α for the fitted power law functions.

|     | Count  | Min | Max    | Median | Geom-Mean | α   |
|-----|--------|-----|--------|--------|-----------|-----|
| AES | 21,415 | 1   | 12,473 | 2      | 2.4       | 2.0 |
| POS | 9,623  | 1   | 186    | 1      | 1.5       | 2.5 |
| TW  | 43,064 | 1   | 8,626  | 3      | 3.2       | 2.1 |

Table 6. Spearman correlation with zero imputed metrics (n=61,848)

|     | AES | POS  | TW   |
|-----|-----|------|------|
| AES | 1   | 0.48 | 0.45 |
| POS |     | 1    | 0.36 |

The data underlying all three metrics is highly skewed—an effect observed and described by various scholars in bibliometrics, see Lotka's and Price's laws (Lotka, 1926; Nicholls, 1988), as well as other altmetric studies (Eysenbach, 2011; Haustein, Peters, Sugimoto, Thelwall, & Larivière, 2014). Accordingly, the direct comparison of the arithmetic mean and variance based on the observed data is limited in its use (Newman, 2005). Instead, several metrics based on the fitting of power law functions and required base parameters (i.e., α and $x_{min}$) have been proposed. Milojević (2010) built on Newman's work to propose a pragmatic approach to estimate the slope α of empirical datasets. Most datasets encountered in research exhibit small sample sizes and considerable noise, which affects the validity and usefulness of fitting theoretical functions as suggested by Newman (2005). Milojević suggests *partial logarithmic binning* of observed values, followed by a simple *least square fit* (LSF) of the newly binned values to determine α. This approach provides an intuitive and effective approach to compare several distributions with the added benefit of a visual representation of the complete data which is not distorted by the noise in the long tail.

---

7 While the reasons for the absence of values varies by collection method (e.g., a page not indexed by Altmetric for POS or a share made without a link for AES), the end result is that, if the method did not find any shares, it is as if they did not happen. Zero imputation applied to 40,433 (65.4%) articles for AES, 52,225 (84.4%) for POS, and 18,784 (30.4%) for TW.

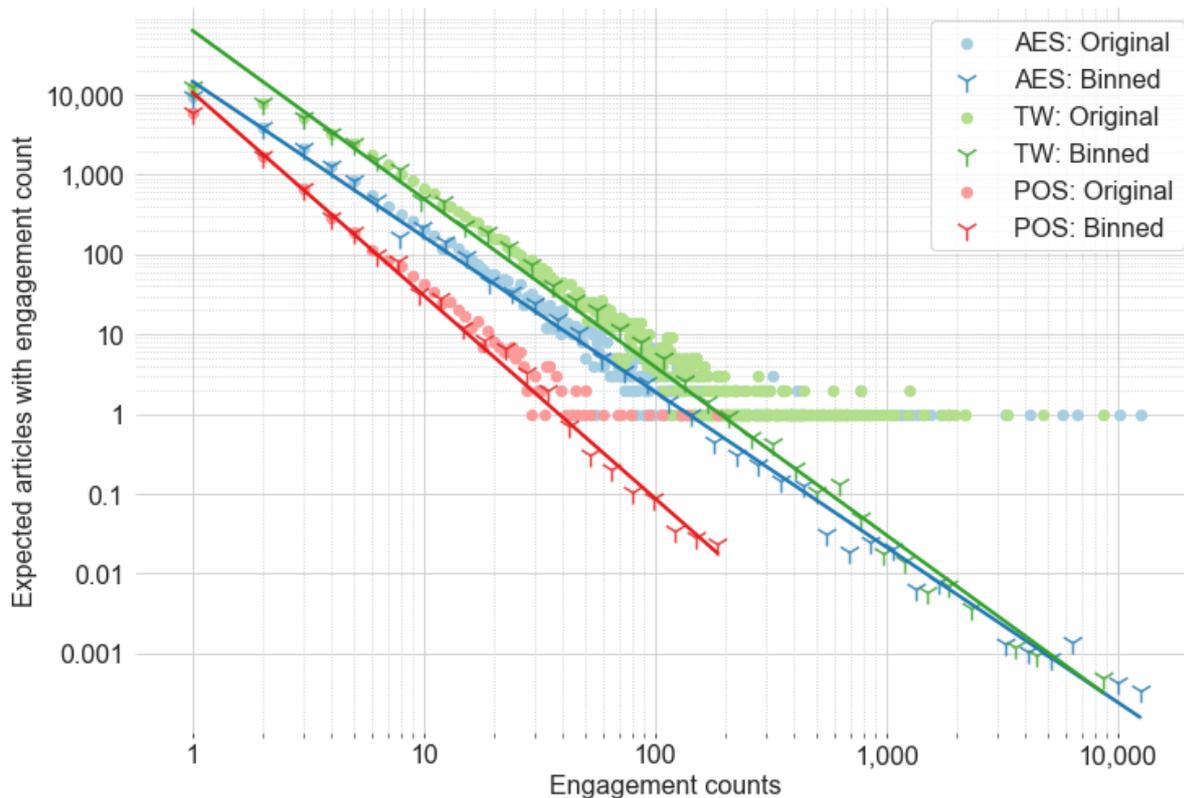

Figure 5. Expected number of articles for observed engagement counts for Twitter (TW), all-engagement shares (AES), and public-only shares (POS). Method of least squares was used on binned values to determine the α of each distribution. I.e., we can read an expected value of roughly 0.01 articles with 1,000 shares, which can be understood as 1 in 1,000 articles with a share count between 891 and 1,122 (the limits for the logarithmic bins are 2.95-3.05).

Using Milojević's (2010) recommended parameters, we used a binning threshold ($k$) *of* 5 and a bin size of 0.11. After fitting LSF slopes (Figure 5), AES shares display the lowest α=2.0, followed by Twitter α=2.1, while POS exhibit the highest one of α=2.5 (i.e., the values taper off quicker for POS in contrast to AES and TW). Furthermore, the original data (especially for Twitter) deviates from the fitted line for lower engagement counts, indicating a hooked power-law distribution, which is another commonly observed and studied property of distributions in the information sciences (Newman, 2005; Thelwall & Wilson, 2014).

Facebook counts in detail

Given that the API (AES) and data mining (POS) methods are both covering the same data source, we can also compare differences in counts for the 7,716 articles that had non-zero counts for both. Of these, 5,223 (67.7%) had more AES than POS, 2,027 (26.3%) articles returned the same number of shares for both methods, while only 644 (6.0%) of all articles returned a higher POS count. Figure 6 shows a letter-value plot[8] for those articles that displayed a difference between counts. Plotting the absolute differences on a logarithmic scale provides a comparison of the medians and several other percentiles. Two observations for articles with a higher POS count stand out: (1) from 2015 to 2017, the number of articles with higher POS than AES counts reduces from 200 to 95 articles and, (2) while the median difference remains stable at 2, the number of articles with higher differences drastically reduces (i.e., the higher percentiles in the plot disappear for POS > AES).

---

8 The letter value plot extends the classic box-plot to provide more detail for big, heavy-tailed datasets. Each bound of a box represents a percentile of the dataset. The lowest box resembles a traditional box-plot with a solid line for the median and the bounds at 1-¼ (75% below the boundaries of the box). Each box then continues to indicate the 1-⅛ (87.5%), 1-1/16 (93.75%), and so on until a border condition is met. Outliers are finally those data points that lie outside of the final plotted box. See Hofmann, Wickham, & Kafadar (2017) for more details about the graph.

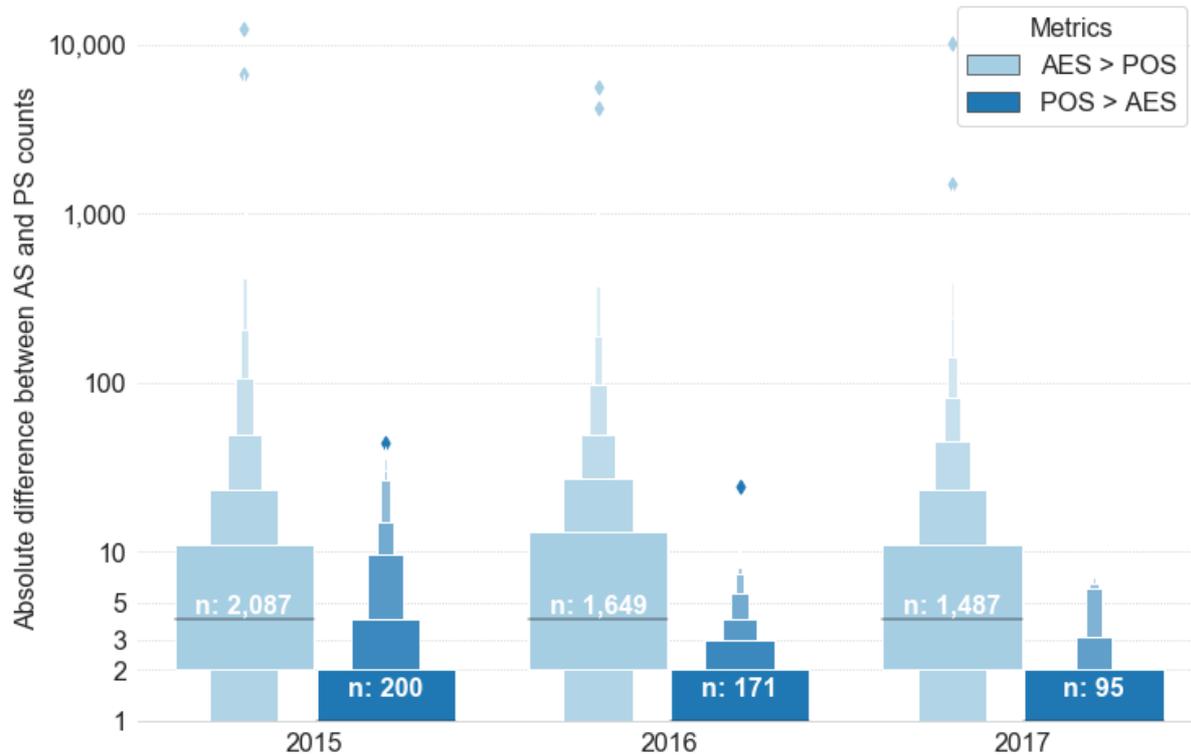

Figure 6: Letter-value-plot of the absolute difference between AES and POS counts across the years.

Facebook counts by discipline

While (as reported above), 67.7% of the articles have a higher AES count, we observe large differences by discipline (Figure 7). In Professional Fields, this ratio even goes up to almost 90%. The ratio drops to 60% for both Clinical Medicine and Biomedical Research—the disciplines with the most publications. With the decreasing share of articles with higher AES counts (from around 90% in Professional Fields down to 60% in Biomedical Research), we observe that the share of articles with equal AES and POS counts increases—from around 10% in the Social Sciences to a third (32.5%) in Biomedical Research. Conversely, the number of articles with higher POS counts never surpasses 7.5% (Clinical Medicine). No articles with higher POS counts were found for Mathematics or Chemistry.

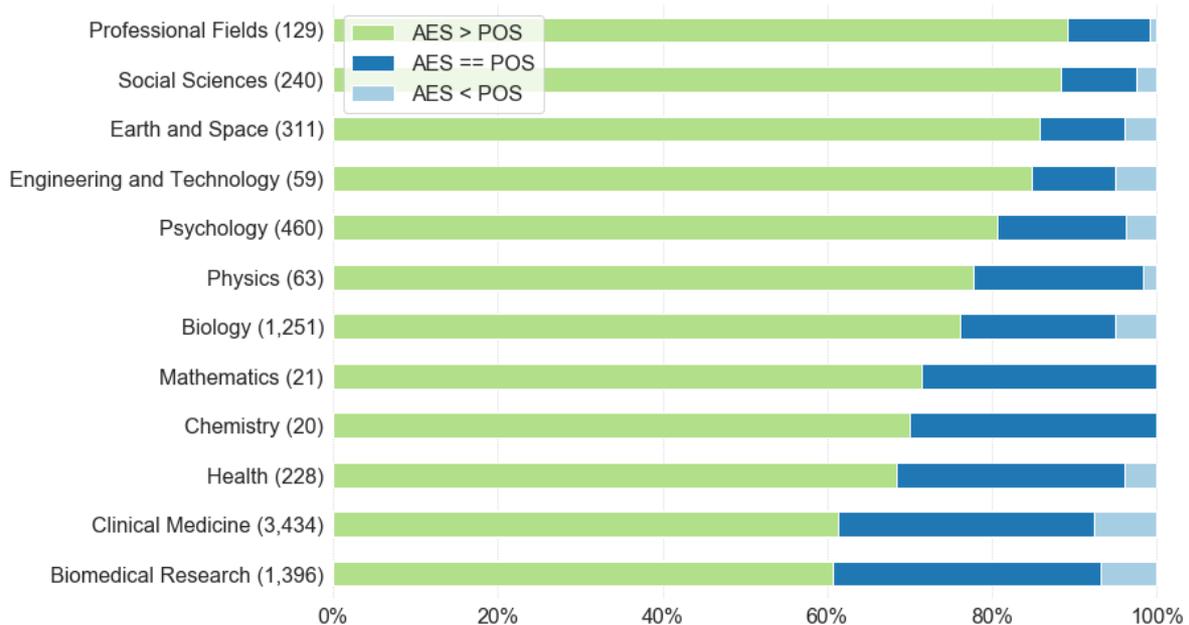

Figure 7: The ratio of articles with more, less, or equal AES and POS counts per discipline.

## 5.0 Discussion and Conclusion

This study compared two data collection methods for retrieving Facebook activity to scientific journal articles. Based on PLOS ONE articles published between 2015 and 2017, we analysed the variations in coverage and extent of the difference in Facebook engagement counts of public Facebook posts retrieved by Altmetric via data mining techniques (POS) with all Facebook engagement counts retrieved via the URL-based Facebook Graph API (AES). In doing so, we found differences in the volume and disciplinary distribution of Facebook engagements captured by both approaches, which serves to highlight that each differs in the *type* and *scope* of engagement they collect. As such, our study serves to highlight the importance of considering which approach is most appropriate in any given context.

The lack of strong correlation between counts found between both methods (Table 6), the different slopes in how engagements of each are distributed (Figure 5), and large disciplinary differences (Figure 7) all show that the activities captured by each method are distinct. Even though POS are largely a subset of AES, the counts derived from each method should be considered distinct indicators. Accordingly, the use and application of either metric should consider how the type and scope of engagement measured by each method relates to the phenomenon of interests. For example, some might argue that the POS method is more appropriate for capturing science communication activities, while the AES method might be more appropriate for understanding individuals' sharing practices. In this study, we remain agnostic to what each metric is best suited for and instead focus on describing how they differ.

We should note that our study benefited from the use of a particularly homogenous dataset from PLOS ONE. The fact that all articles were published on a single website facilitated the identification of relevant URLs to be queried in the Facebook Graph API and allowed us to be more precise in our data collection and comparisons than would have otherwise been possible. Studies seeking to use the AES approach across publishers may need to consider using different URL variants for each publisher website and for the multiple places in which articles may appear (e.g., if they are re-published by an aggregator). This limitation may also hold true for the POS approach, but it is particularly evident in the AES approach. This said, efforts are underway to implement the AES approach on a large scale (Alperin, Enkhbayar, Piwowar, Priem, & Wass, 2018). Moreover, our code is available for others to experiment (Enkhbayar, 2019).

Our analysis shows that, at least under these idealized conditions, more than half of all Facebook engagement takes place between users and therefore is not counted by the POS approach. This shows that working with public-only pages captures only a subset of the research-related activity that happens on the platform. The large share of activity in user's Timelines seems to corroborate findings by McClain (2017), who shows that academics on Facebook tend to share science on the platform in a personal and intimate manner, rather than professionally for science outreach and communication.

Although AES coverage values for PLOS ONE papers are still only half that of Twitter (34.6% vs. 69.6%), they are also twice as much as what was found with POS (15.6%), and roughly double of what was previously reported for PLOS ONE publications (Zahedi & Costas, 2018b). This suggests that the importance of Facebook as a data source for researchers interested in science communication and public engagement on social media may have been underestimated by altmetrics researchers. The proportion of articles covered across individual years remains stable for each method, even as the number of social media users still increases annually, albeit more slowly on Twitter (Facebook, 2019; Twitter, 2019). This might indicate a saturation of the Facebook and Twitter communities with regard to sharing scientific research articles.

Coverage varied widely by discipline for each metric. However, disciplines usually ranked similarly for each metric in terms of coverage indicating that disciplinary differences applied to all data sources. Almost all Psychology articles were tweeted (94.2%), while Chemistry had the lowest Twitter coverage (44.4%). This lower value is still higher than the highest POS coverage (Social Sciences at 35.0%), but lower than the highest AES coverage (Social Sciences at 70.5%).

To better understand the differences between AES and POS methods we compared articles that were indexed by only one of the two approaches with each other. One third of articles were found by both methods, while 58.7% were only retrieved by AES. Likely due to an advantage of the POS approach, which finds references to a wider range of URLs across scholarly domains, Altmetric is able to retrieve 8.2% additional articles with Facebook activity, which were not discovered the URL-based AES approach.

More importantly, the difference in coverage between the POS and AES methods varies across disciplines, putting the so-called hard sciences (i.e., Engineering and Technology, Chemistry, Physics, Mathematics) at a greater disadvantage when measured by public posts. For these disciplines, AES found three to four times the number of articles POS did. Even in disciplines where the advantage was lower (e.g., Social Sciences, Health, Psychology, and Professional Fields) the AES approach still produced twice as many results.

This proportionally lower ranking of the hard sciences by Altmetric's implementation of the POS approach has important implications for all altmetrics studies using their data. For example, Haustein, Costas and Larivière (2015) report the lowest Facebook coverages for these fields but would likely have yielded different results if using the AES method. Part of this lower coverage may thus not be due to lower engagement overall, but rather to the nature of that engagement. On the other end of the spectrum, although still much lower than articles discovered by AES only, the share of documents discovered by only POS is particularly high in Clinical Medicine. These articles might be missed by AES because they are being shared with a different URL (known to Altmetric but not included in the 10 variations used by us). Overall, this, once again, underlines the importance of paying more attention to the *scope of engagement* in altmetric studies involving Facebook data.

We also showed that different retrieval techniques vary in the number of Facebook shares they uncover for each article. As to be expected and shown previously for other citation and social media-based indicators, the distributions of Facebook and Twitter activity per document is skewed (Newman, 2005). Furthermore, like Thelwall and Wilson (2014) report for citations, the results indicated hooked power-law distributions, especially clearly for Twitter. It is further interesting to note that Facebook counts that include both public and private engagements resemble public Facebook counts only moderately ($\rho=.48$), at the same level as they resemble tweets ($\rho=.45$), while correlations between POS and TW are lower ($\rho=.36$). The analysis of power-law distributions and correlations suggest that public and private engagement on Facebook related to scholarly articles are not identical. Furthermore, Facebook engagement that includes private spaces seems to more closely resemble tweeting behavior than public engagement on Facebook does. Thus, the collection of public engagement data should not be used to predict private engagement counts.

Among the 7,715 PLOS ONE papers retrieved by both methods, engagement counts were identical for around one quarter (26.3%) of articles. AES counts exceeded PLOS in two thirds (67.7%) of the cases, while POS reported higher counts than AES 6.0% of the time. Over the three years analyzed, the share as well as the extent to which POS exceed AES counts continuously decreased. On the disciplinary level, results vary dramatically across fields. In the Professional Fields AES counts exceeded POS for almost 90% of the articles, which demonstrates a particular strong tendency by users to share scientific articles in private spaces. On the other end of the spectrum, Clinical Medicine and Biomedical Research had a larger share of public Facebook engagement, even if more than half of all publications' engagement still took place outside of public pages and groups. The proportion of articles with equal POS and AES counts also varies, with disciplines that have a higher number of matches tending to have a lower proportion of articles with higher AES counts drops. That is, at the discipline level, AES counts are either higher than POS counts, or they are equal. As a result, the number of articles with higher POS counts never surpasses 7.5% (the proportion reached by Clinical Medicine).

Our study shows that the retrieval method has a significant effect on results and commonly used altmetrics. Findings demonstrate that reliability and reproducibility of Facebook data in the context of altmetrics is questionable (Haustein, 2016). These differences have far-reaching implications for our understanding of the role that Facebook plays in the diffusion of and engagement with scholarly articles by highlighting the significantly higher levels of engagement that take place outside of public spaces on the platform. We therefore urge all altmetric data aggregators and authors of altmetric studies to carefully consider the type and scope of engagement that is relevant and to be transparent and open about the shortcomings of respective retrieval methods.

## Declaration of Conflicting Interests

## Note on Research Ethics

This study is part of a larger project and is registered at the Office of Research Ethics of Simon Fraser University with the Study Number 2016s0170. Despite reporting on data about engagements that happen outside of public view, this study relies entirely on publicly available aggregate data. It did not, as the methods describe, do any data collection requiring the use of private tokens for access. That is to say, the data themselves are not actually "private"—what is private are the posts in which some of the sharing took place and contains no information that might be used to identify individual subjects. What is more, the analysis is primarily about the nature of links to scientific articles shared on Facebook, and, as such, does not qualify as human subject research. This was the understanding of the Office of Research Ethics at Simon Fraser University where we registered our study and is supported by both commonly used definitions of human subject research as the one provided by the NIH[9] and by more specific web research ethics frameworks[10].

## Acknowledgements

We would like to thank and acknowledge Altmetric for access to the Facebook (POS) and Twitter data, as well as for comments, feedback, and clarifications on earlier version of this manuscript. We also acknowledge the Observatoire des sciences et des technologies (OST), Centre interuniversitaire de recherche sur la science et des technologies (CIRST) for access to their in-house database including the Web of Science and our data. Finally, we would like to thank Joe Wass (Crossref) and Scott Chamberlain (rOpenSci) for detailed technical conversations and writings on the URLs, DOIs, and APIs. Our development of the AES approach would not have been possible without their work.

## Funding

This study was supported by the Social Sciences and Humanities Research Council of Canada through Grant (892-2017-2019) to JPA and SH. The funders had no role in study design, data collection and analysis, decision to publish, or preparation of the manuscript.

---

9 https://grants.nih.gov/policy/humansubjects/research.htm
10 Bowser, A., & Tsai, J. Y. (2015, May). Supporting ethical web research: A new research ethics review. In *Proceedings of the 24th international conference on world wide web* (pp. 151-161). doi: 10/dpct

## Authorship Contributions

*Asura Enkhbayar.* ORCID: 0000-0002-3934-026X. Conceptualization, Data curation, Formal Analysis, Investigation, Methodology, Software, Visualization, Writing – original draft, Writing – review & editing

*Stefanie Haustein*. ORCID: 0000-0003-0157-1430. Formal Analysis, Funding acquisition, Investigation, Writing – review & editing

*Germana Barata.* ORCID: 0000-0001-6064-6952. Conceptualization, Investigation, Writing – review & editing

*Juan Pablo Alperin*. ORCID: 0000-0002-9344-7439. Conceptualization, Formal Analysis, Funding acquisition, Investigation, Methodology, Project administration, Supervision, Writing – original draft, Writing – review & editing

## Data Availability

The data used to produce results is available at doi:10.7910/DVN/3CS5ES (Enkhbayar, Haustein, & Alperin, 2019) and has been published under a CC0 license. The code to reproduce all figures and tables is available at doi:10.5281/zenodo.3381821 (Enkhbayar, 2019) and is published under the MIT license. The code to collect the original data can be found at doi:10.5281/zenodo.1314990.

## Appendix A

| Discipline | AES (coverage) | POS (coverage) | TW (coverage) | All articles (percentage) |
|---|---|---|---|---|
| Clinical Medicine | 9,668 (32.9%) | 4,630 (15.8%) | 20,387 (69.4%) | 29,360 (50.7%) |
| Biomedical Research | 4,554 (31.7%) | 1,760 (12.3%) | 10,379 (72.3%) | 14,351 (24.8%) |
| Biology | 3,069 (45.4%) | 1,355 (20.0%) | 5,150 (76.2%) | 6,761 (11.7%) |
| Psychology | 1,080 (64.2%) | 515 (30.6%) | 1,583 (94.2%) | 1,681 (2.9%) |
| Engineering and Technology | 314 (26.5%) | 78 (6.6%) | 634 (53.6%) | 1,183 (2.0%) |
| Earth and Space | 593 (54.0%) | 327 (29.8%) | 802 (73.0%) | 1,098 (1.9%) |
| Health | 567 (52.6%) | 276 (25.6%) | 893 (82.9%) | 1,077 (1.9%) |
| Social Sciences | 516 (70.5%) | 256 (35.0%) | 646 (88.3%) | 732 (1.3%) |
| Physics | 261 (42.4%) | 73 (11.9%) | 424 (68.8%) | 616 (1.1%) |
| Professional Fields | 297 (62.7%) | 141 (29.7%) | 416 (87.8%) | 474 (0.8%) |
| Chemistry | 83 (24.6%) | 27 (8.0%) | 150 (44.4%) | 338 (0.6%) |
| Mathematics | 74 (32.0%) | 26 (11.3%) | 144 (62.3%) | 231 (0.4%) |
| *Total* | *21,076 (36.4%)* | *9,464 (16.3%)* | *41,608 (71.9%)* | *57,902 (100.0%)* |

Table A1. Numeric values of the coverage of disciplines for AES, POS, and TW as displayed in Figure 3 including the proportion of all articles.

# Appendix B

| Discipline | Any FB | Only AES | Both | Only POS |
|---|---|---|---|---|
| Clinical Medicine | 10864 | 6,234.0 (57.4%) | 3,434.0 (31.6%) | 1,196.0 (11.0%) |
| Biomedical Research | 4918 | 3,158 (64.2%) | 1,396 (28.4%) | 364 (7.4%) |
| Biology | 3173 | 1,818 (57.3%) | 1,251 (39.4%) | 104 (3.3%) |
| Psychology | 1135 | 620 (54.6%) | 460 (40.5%) | 55 (4.8%) |
| Health | 615 | 339 (55.1%) | 228 (37.1%) | 48 (7.8%) |
| Earth and Space | 609 | 282 (46.3%) | 311 (51.1%) | 16 (2.6%) |
| Social Sciences | 532 | 276 (51.9%) | 240 (45.1%) | 16 (3.0%) |
| Engineering and Technology | 333 | 255 (76.6%) | 59 (17.7%) | 19 (5.7%) |
| Physics | 271 | 198 (73.1%) | 63 (23.2%) | 10 (3.7%) |
| Professional Fields | 309 | 168 (54.4%) | 129 (41.7%) | 12 (3.9%) |
| Chemistry | 90 | 63 (70.0%) | 20 (22.2%) | 7 (7.8%) |
| Mathematics | 79 | 53 (67.1%) | 21 (26.6%) | 5 (6.3%) |
| *Total* | *22928* | *13,464 (58.7%)* | *7,612 (33.2%)* | *1,852 (8.1%)* |

Table B1. Proportion of articles found by the two different Facebook methods broken down into: Found by AES only (private posts), POS only (public posts), and by both methods. This table reflects the data in figure 4.

# Supplementary Material

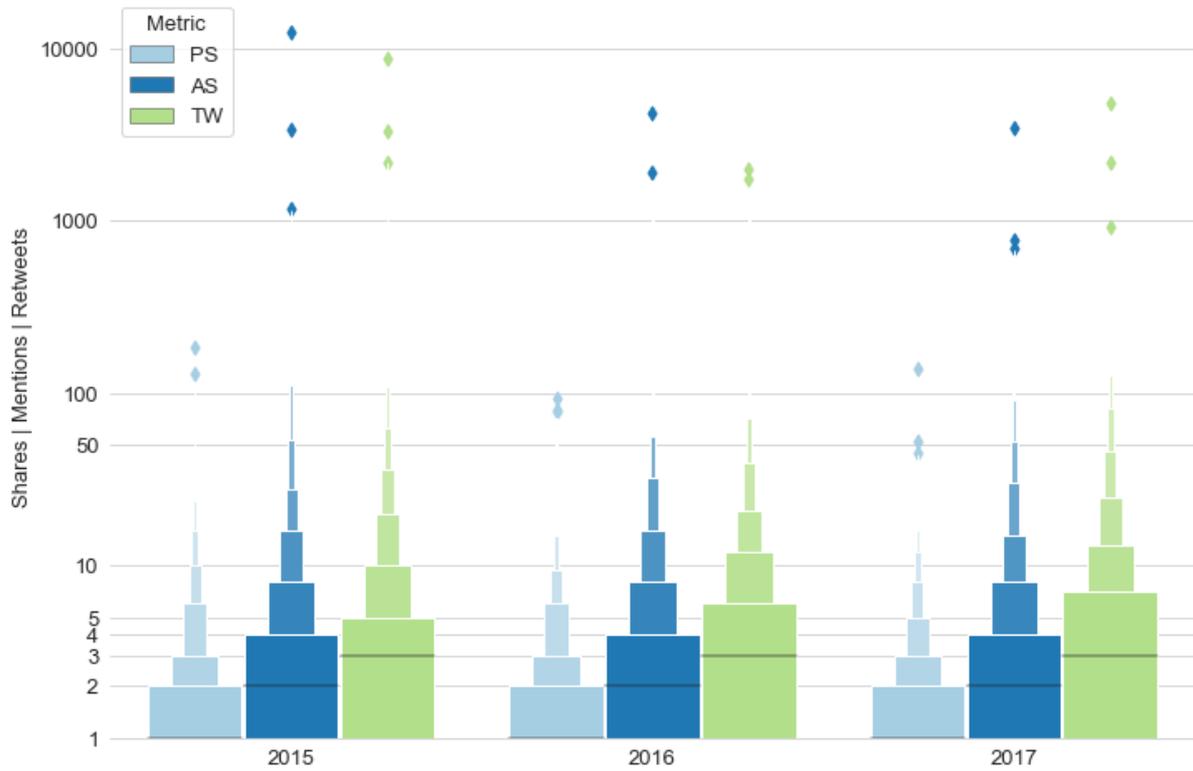

Figure 8. Letter-Value-Plot for POS, AES, and TW across the years. The three distributions remain mostly constant. The percentiles at 50%, 75%, and 87.5% remain constant for AES and POS over all years. While the data in the tails is displaying minor changes, the overall engagement counts remain stable. Only Twitter shows increases beyond the 75th percentile indicating a trend towards higher engagement counts.

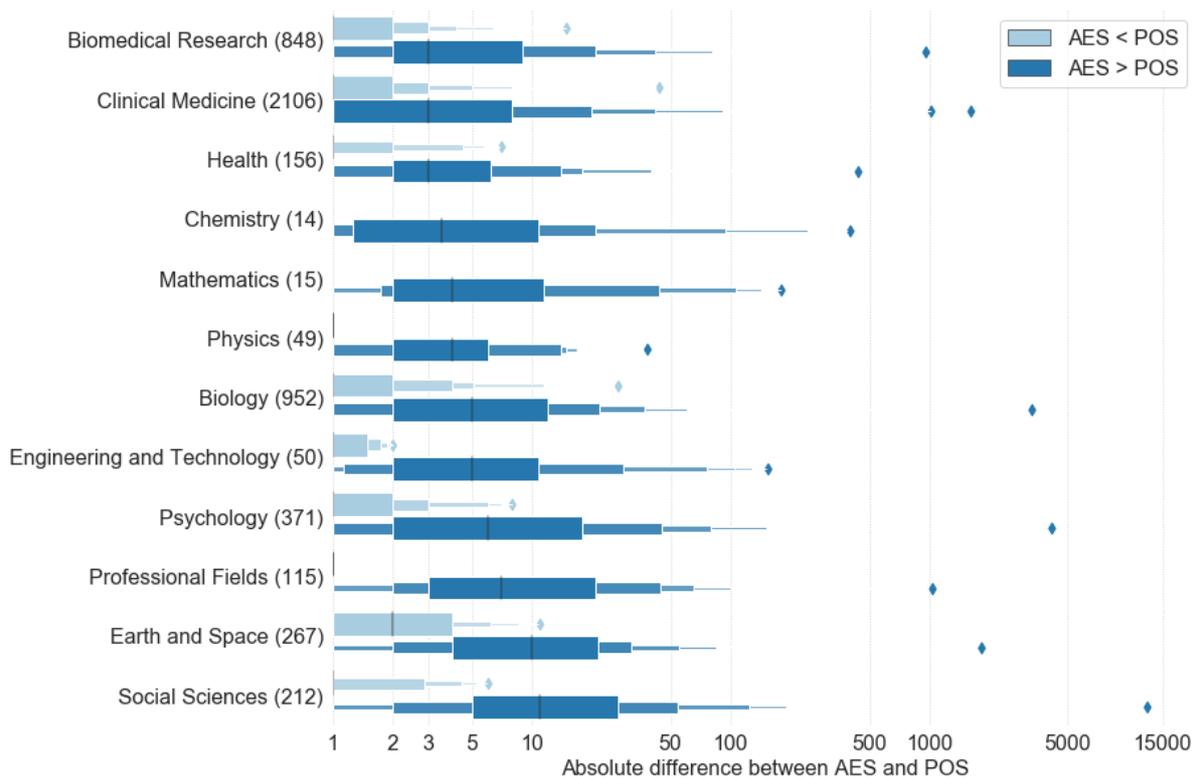

Figure 9. Letter-value-plot of the absolute difference between AES and POS counts. The lowest median difference between private and public shares canhas been observed for Biological Medicine, Clinical Medicine, and Health with 3 shares. The disciplines with the highest median difference are Earth and Space and Social Sciences, 10 and 11 shares respectively. The remaining disciplines all display a median difference of 4 to 7 shares. It is furthermore notable that the median difference for articles that had higher POS counts is 1 for all disciplines but Earth and Space, which reports a median difference of 2.